# Simulation of Single-Phase Natural Circulation within the BEPU Framework: Sketching Scaling Uncertainty Principle by Multi-Scale CFD Approaches

## Haifu Huang[1, 2], Jorge Perez[1*], Nicolas Alpy[1], Marc Medale[2]

[1] French Alternative and Atomic Energies Commission - CEA, IRESNE/DES/DER/SESI, 13108 Saint Paul lez Durance, France

[2] Aix-Marseille University - IUSTI, UMR 7343 CNRS, Technopôle de Château-Gombert, Marseille, France

## ABSTRACT

For safety enhancement purposes, nuclear reactors consider natural circulation mechanisms as a mean to remove residual power in its design phase. The simulation of this passive mechanism must be qualified when transitioning from the validation range to the scope of utilization (reactor case), introducing potential physical and numerical distortion effects. In this study, we simulate the flow of liquid sodium using the TrioCFD code, employing both higher-fidelity (HF) LES and lower-fidelity (LF) URANS models. We tackle respectively numerical uncertainties through the Grid Convergence Index method, and physical modelling uncertainties through the Polynomial Chaos Expansion method available on the URANIE platform. HF simulations exhibit a strong resilience to physical distortion effects, with numerical uncertainties being intricately correlated. Conversely, the LF approach, the only one applicable at the reactor scale, is likely to present a reduced predictability. If so, the HF approach should be effective in pinpointing the LF weaknesses: the concept of scaling uncertainty is inline introduced as the growth of the LF simulation uncertainty associated with distortion effects. Thus, the paper outlines that a specific methodology within the BEPU framework - leveraging both HF and LF approaches - could pragmatically enable correlating distortion effects with scaling uncertainty, thereby providing a metric principle.

## 1. INTRODUCTION

Passive systems have become an important asset in nuclear safety reducing dependencies on active controls, external power supply or operator interventions. Systems using natural circulation are integrated across various reactor concepts, significantly increasing safety and reliability [1]. However, the magnitude of the natural forces, which drive the operation of passive systems, is relatively small and counterforces (e.g., friction) cannot be ignored for flow onset, as it is generally the case of systems including pumps. Additionally, possible cliff edge effects (such as the existence of a threshold for convection onset or corresponding to a dynamic instability around a steady state [2]) must be identified, mapped and quantified so that a safe operation can be designed. Due to the complexity of their physical mechanisms, a primary challenge lies in their scaling. Actually, scaling is not a new issue for nuclear safety licensing. The challenges come from the inability to conduct reactor-scale tests, the difficulties in achieving exact physical and operational parallels in down-scaled models, and the complexities involved in accurately qualifying the up-scaling processes. As highlighted in the OECD/NEA report [3], scaling uncertainties arise mainly from experimental limitations in reproducing safety related transient physics without simplifications, both under design and experimental conditions, and from inherent limitations of numerical tools for solving the exact equations.

Over decades of R&D, the scaling process has been considered into the evolving Best Estimate Plus Uncertainty (BEPU) framework as part of safety licensing. Early works on the scaling investigation have highlighted two mainstream methodologies. The Code Scaling Applicability and Uncertainty (CSAU) methodology proposes the evaluation of up-scaling capacity on the test facility and applied codes, where the correlations of uncertainties show the possibility of scaling uncertainty quantification [4]. Consistent with CSAU, the Uncertainty Methodology based on Accuracy Extrapolation (UMAE) methodology, which relies on an extrapolation method, offers a four-step strategy to address the scaling issue [5]. In line with the requirements of the French Nuclear Safety Authority (ASN) in Guide 28, it is crucial to specify how the conclusions of Scientific Computing Tool (SCT) validation apply to the intended scope of utilisation [6]. The demand is logically reinforced as advancement in computational capabilities (especially for single-phase flow), opens new possibilities regarding predictability but raise also new methodological challenges. The complexities of scaling for nuclear safety licensing, particularly in the context of natural circulation within passive systems, require meticulous attention when being assessed.

In order to investigate the scaling issue for a Natural Circulation Loop (NCL) under BEPU framework, different fidelity simulations have been leveraged to evaluate the scaling uncertainties. A Horizontal Heating Horizontal Cooling (HHHC) loop case has been selected as a configuration allowing a two-sided methodological advance for scaling issues: on the threshold for convection onset (see [12]) and on flow steady-state. Simulations are conducted using TrioCFD code[1], incorporating specific physical and numerical modelling choices. Numerical and physical modelling uncertainties are also quantified. Finally, uncertainty is assessed with respect to specific distortions so that the scaling effect can be analysed.

## 2. CFD SIMULATIONS ACCORDING TO DIFFERENT FIDELITY APPROACHES

In this section, we outline some specifics of the considered NCL which simulation requires physical models and numerical methods distinct from those of forced convection due to some peculiar nonlinear physics. The different simulation approaches carried-out with TrioCFD are presented as well as the applied guideline for NCL computations under BEPU framework.

### 2.1 Physical phenomena of a NCL with liquid sodium coolant

The considered HHHC NCL, which is extracted from IAEA literature, is sketched in Figure 1 (left). While pressure-driven flow corresponding to forced convection via pumps have been extensively studied, natural circulation using liquid sodium presents unique challenges for CFD simulations. These challenges include considering coupled impacts on the development of bulk flow and boundary layers from boundary conditions (e.g. flux or temperature imposed for the heat source and sink), specific properties of sodium, buoyancy forces and curvature effects due to bends. As the flow develops, we observe anisotropic turbulence phenomena, such as Dean vortices in Figure 1 (right) which significantly influence the pressure drop and possibly heat transfer depending on bends location. Moreover, for low Prandtl number flow like liquid sodium

---

[1] https://triocfd.cea.fr/ on TRUST platform https://github.com/cea-trust-platform

($\sim 7.10^{-3}$), the large difference between hydraulic and thermal boundary layers requires wall-resolution, as closure laws for wall-modelling are found unreliable. The symmetrical nature of the loop makes the onset of natural circulation very specific (pitchfork bifurcation), necessitating sufficient numerical noise for activation, therefore adding another non-linear physics challenge.

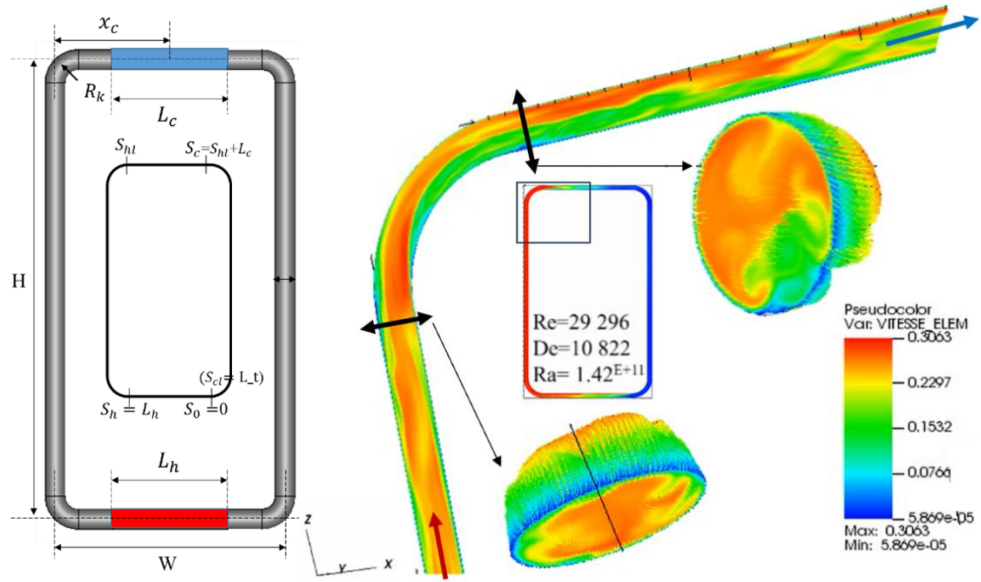

**Figure 1: 3D-velocity field in the HHHC NCL.**

## 2.2  Physical and numerical modelling by TrioCFD

The TrioCFD, CEA's in-house code, is an extension of the open-source TRUST platform. It is equipped to handle various types of equations, offering a wide range of time schemes, spatial discretizations, mathematical solvers, and boundary conditions [7]. In our study, we employed two turbulence modelling approaches: Large Eddy Simulation (LES), which includes a WALE subgrid model, and Reynolds Averaged Navier-Stokes (RANS). Two RANS approaches were investigated: a standard linear κ-ε model (Linear Eddy-Viscosity Model, LEVM) with wall laws and a wall-resolved Non-Linear Eddy-Viscosity Model (NLEVM) proposed by Baglietto [8]. Despite some limitations in physical representation, 2D simulations were primarily chosen for their efficiency and lower CPU demands, focusing first and foremost on methodological development, while still capturing crucial phenomena like centrifugal and buoyancy forces.

The CFD study of NCL requires unstationary approaches like DNS, LES or URANS to gain access to fluctuation intensity and frequency. Specifically, LES is particularly advantageous in NCL studies as it captures the largest turbulent structures, with small-scale motions represented by a subgrid-scale (SGS) model. This approach is preferred for accurately depicting velocity distributions, especially downstream of elbows, where secondary flows are significant. Figure 2a shows a wall-resolved LES in 2D, highlighting how Dean Vortices lead to local recirculations at the elbow downstream, impacting the boundary layer and simulation quality. In addition, with mesh refining from M1 to M3, the averaged velocities at half-height are compared. Notably, LES solutions tend to converge more effectively than 'DNS' (with same meshes) as shown in Figure 2b, offering a pragmatic mean to reduce CPU cost while achieving high fidelity computation.

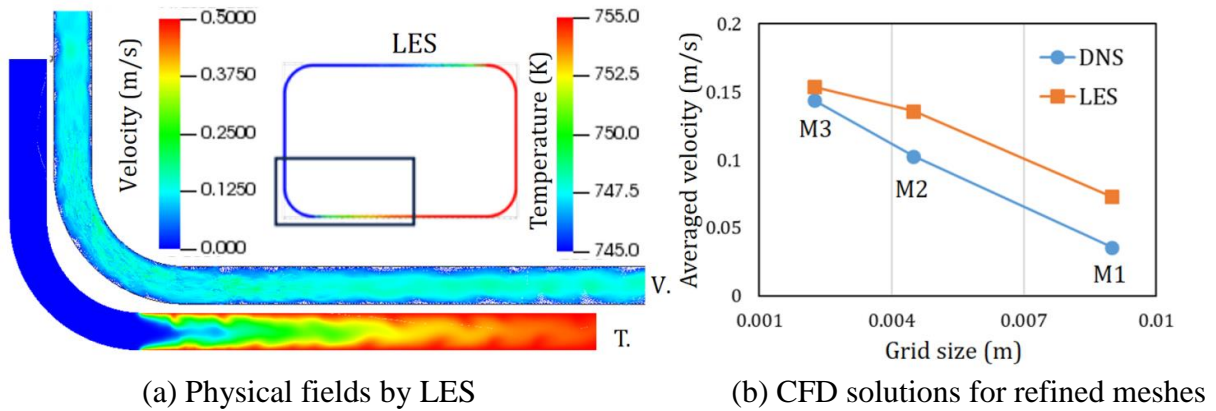

(a) Physical fields by LES  (b) CFD solutions for refined meshes

**Figure 2：2D-physical fields and solution grid convergence.**

Figure 3 visually contrasts the performance of two turbulence models in capturing local recirculation due to centrifugal effects in a NCL flow. The linear LEVM, using a classical wall function approach, generalizes the details in both velocity and temperature fields, thus failing to capture the complexities of anisotropic turbulence. In contrast, the NLEVM with low Reynolds treatment near walls extends Reynolds stresses non-linearly, enhancing anisotropy representation, making it better suited for scenarios with significant recirculation. It effectively demonstrates the downstream propagation of local recirculation and reveals extended thermal mixing.

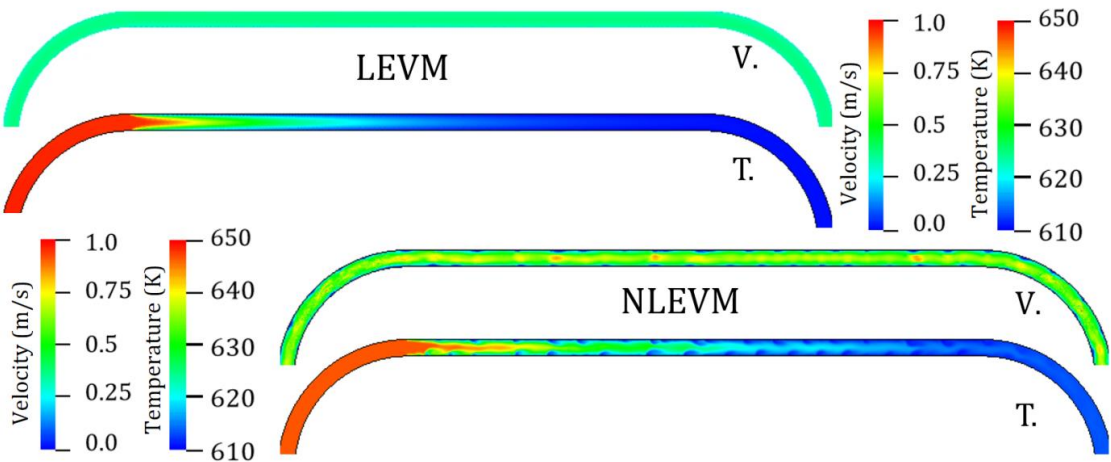

**Figure 3: 2D-physical fields for eddy-viscosity models with different wall treatments.**

Figure 4 displays a comparative analysis of instantaneous velocity profiles across the flow section, captured after reaching a statistically steady state. This data is captured downstream one of the elbows (where the centrifugal effects are most pronounced), using various turbulence models. For a phenomenon-focused objective, compared to the wall-resolved simulations (case 2 & 3 with $y^+$ ranging from 0.5 to 8), the wall-modelled ones (case 1, 4 & 5 with $y^+$ around 30) cannot capture intricate fluid behaviours, such as flow reversal in local recirculation near the wall. Especially, wall-resolved LES (case 2) effectively captures near-wall natural circulation physics, but an improper wall function with LES (case 1) can distort near-wall velocity profiles and bulk flow.

However, for objectives centred on parameters of interest, like flow rate magnitude, strategically adjusting the wall function (specifically, Von Karman coefficient (k) within the logarithmic law for wall functions) enables us to approximate the average flow rate to the LES results when substantially reducing CPU cost. Given that Von Karman coefficient traditionally stands at 0.415 for pressure-driven flows and varies for natural convection, this method of physical model calibration offers a pragmatic trade-off between CPU cost and model accuracy. However, reliability of such a calibration for another scale may arise some concerns, therefore this methodology can be considered as a research domain where further investigate transposition and scaling uncertainty topics.

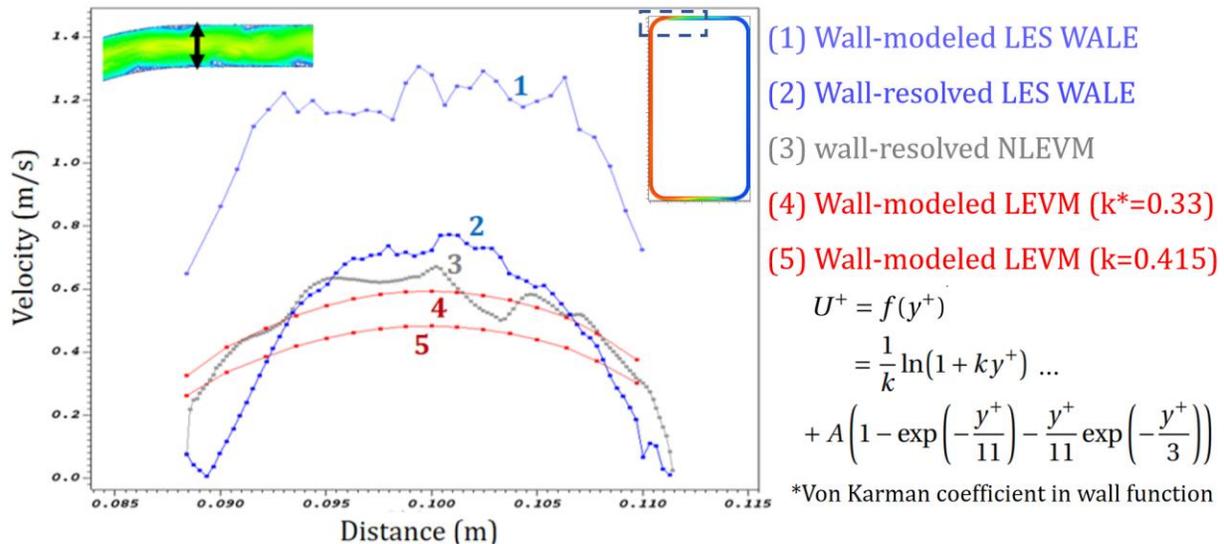

**Figure 4: Radial velocity profiles downstream of an elbow for different physical models.**

### 2.3 CFD guideline, applied for the NCL modelling

The previous section illustrated numerical and physical complexities to be addressed for NCL simulations. Table 1 provides the comprehensive guideline applied for NCL computations under BEPU framework [11], addressing phenomena identification, modelling and simulation while using TrioCFD. Verification and validation of the simulation solution are also part of the qualification. Verification process can be divided into code and solution verification following ASME standard [9]. Especially, the solution verification involves errors estimation for the unknown exact solution by using Richardson extrapolation and its uncertainty quantification by applying Roache's Grid Convergence Index (GCI) [10]. Then for the validation process specified in ASN guideline, both separated and integral validation should be investigated against the physical phenomena [6]. This comprehensive qualification panel should be extended to scaling effects on modelling uncertainties across different fidelity levels. Ultimately, Figures-of-Merit (FoM) and related uncertainties can be quantified.

**Table 1: CFD guideline for NCL computations according to different fidelity levels.**

| Process | | | Higher-Fidelity | Lower-Fidelity |
|---|---|---|---|---|
| Physical Phenomena | PIRT | | NCLs Physical Behaviors | |
| Physical Modelling | Physical properties | | Liquid sodium | |
| | Assumptions and Approximations | | Compressibility/Boussinesq's approx. | |
| | Turbulence closure law | | LES - WALE | URANS std. $\kappa - \varepsilon$ |
| | Solid Walls Treatment | | Wall resolved | Wall modeled |
| Numerical Modelling | Domain/Mesh/Partition | | METIS library | |
| | Numerical Methods | Discretization technique | Finite Volume Elements (VEF) | |
| | | Temporal discretization | Implicit ($\geq$ 2nd) | Euler implicit |
| | | Spatial discretization | Center/Hybrid | Upwind |
| | | Iterative solvers | PETSc library | |
| Verification | Code Verification | | See ASME standard or TrioCFD | |
| | Solution Verification | | Discretization/Iterative errors … | |
| Validation | Separate effects validation | | Separate Effect Test (SET) | |
| | Integral validation | | Integral Effect Test (IET)/ Integral Test Facility (ITF) | |
| Uncertainties | Numerical solution uncertainties | | Grid Convergence Index (GCI) | |
| | Physical modelling uncertainties | | Polynomial Chaos Expansion (PCE) | |

## 3. QUANTIFICATION OF MODELLING UNCERTAINTIES

In this chapter, we address the quantification of modelling uncertainties. Figure 5 outlines key sources of modelling uncertainties, including the inevitable user effect. We concentrate on two critical uncertainties in CFD simulations: numerical solution uncertainties by GCI method and the physical modelling uncertainties: for the latter, PCE meta-modelling is first carried out by using CEA uncertainty platform URANIE [11].

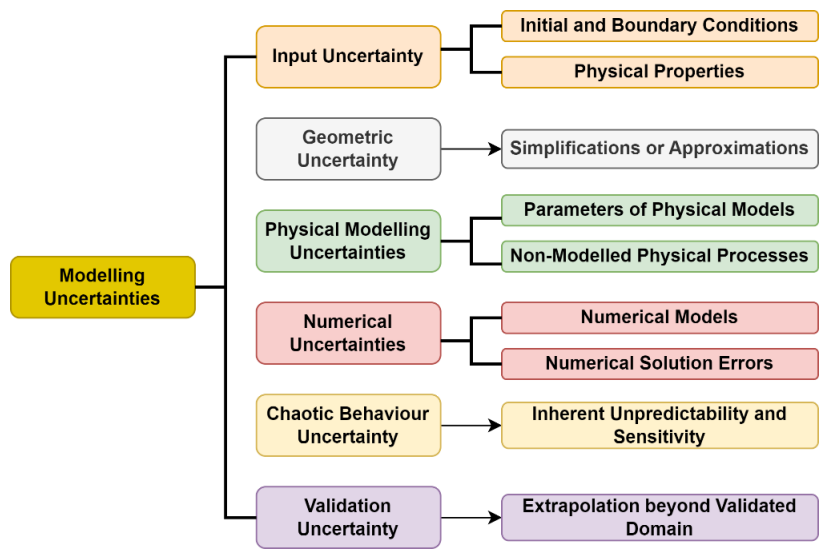

**Figure 5: Potential sources of modelling uncertainties**

### 3.1 Numerical uncertainties by Grid Convergence Index (GCI)

As illustrated in Table 2, the core principle of the GCI method involves using three simulated solutions to determine a converged order $p$. The quality of these solutions determines the safety factor $F_s$ (empirical value given in [10]) applied to convert numerical errors into uncertainties. This approach allows us how to establish an uncertainty band over the finest solution. For instance, Table 3 describes indicative values for the converged order and the corresponding uncertainty band, along with the confidence interval presented over the finest solution. As illustrated in Figure 6, simulations from coarsest $f_1$ (with Mesh 1) to finest $f_3$ (with structured refined Mesh 3) are required to calculate the extrapolated solution $f_{\text{ext}}$ (theoretical value at zero grid size). In addition, the numerical uncertainty is depicted as $CI_{GCI}$ and placed over the finest solution.

**Table 2: Estimation of numerical solution errors and uncertainties through GCI method.**

| | | |
|---|---|---|
| **Numerical solution errors** | Grid size $h_i$ | $h_1 > h_2 > h_3$ |
| | Computed solutions $f_i$ | $f_1; f_2; f_3$ |
| | Refined ratio $r$ | $r = h_2/h_3 (> 1.3)$ |
| | Apparent order $p$ | $p = \ln\left(\frac{f_1 - f_2}{f_3 - f_2}\right)/\ln(r)$ |
| | Extrapolated solution $f_{ext}$ | $f_{ext} = f_3 + \frac{f_3 - f_2}{r^p - 1}$ |
| | Normalized relative error $e_n$ | $e_n = \left|\frac{f_3 - f_2}{f_3}\right|$ |
| **Conversion** | Safety factors $F_s$ | 3 (or 1.25) |
| **Numerical solution uncertainties** | Grid Convergence Index GCI | $GCI = F_s \cdot \frac{e_n}{r^p - 1}$ |
| | Uncertainty band $U_{GCI}$ | $U_{GCI} = GCI \cdot f_3$ |
| | Penalized uncertainty $U_{num}$ | $U_{num} = U_{GCI}/k, k = 1.1{\sim}1.15$ |
| | Confidence Interval $CI_{GCI}$ | $CI_{GCI} = [f_3 - U_{num};\ f_3 + U_{num}]$ |

**Table 3: Indicative values for GCI method application with $F_s$=3**

| $f_i$ | $r$ | $p$ | $f_{ext}$ | $e_n$ | $GCI$ | $U_{num}$ | $CI_{GCI}$ |
|---|---|---|---|---|---|---|---|
| 129.2>11.4.1>61.2 | 2 | 1.81 | 135.2 | 0.117 | 0.14 | 16.43 | [112.76; 145.63] |

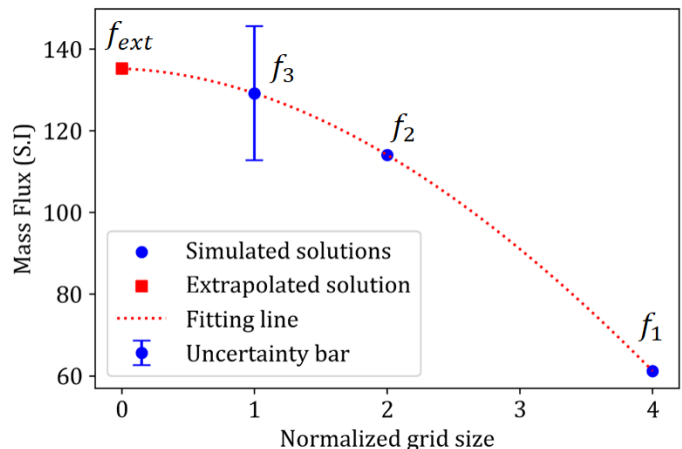

**Figure 6: Illustration of numerical uncertainties through GCI method by LES WALE.**

### 3.2 Physical modelling uncertainties by Polynomial Chaos Expansion (PCE)

To estimate physical modelling uncertainties, we employ the URANIE platform[2] coupled with TrioCFD. Taking the LEVM with wall function as an example, we start by specifying and building our input variables for the physical models. The Von Karman coefficient from the wall law and the turbulent Prandtl number from the energy equation are selected for this analysis. We apply PCE as our meta-modelling approach, with a 4th order polynomial, where a minimum of 15 calculations are needed, as dictated by the cardinality law. Then the design of experiments (DoE) is executed using the classic Latin Hypercube Sampling (LHS) method. By sampling inputs via URANIE and conducting 25 TrioCFD calculations, we compute the PCE coefficients, as shown in the matrix in Figure 7. To qualify a meta-model for the mass flux as quantity of interest, in Table 4 we assess criteria such as $R^2$, $Q^2$ and validation error applying data not previously used for training. Meanwhile, despite more advanced technique applied, the cross-validation with UQLab[3] confirms the effective application of URANIE.

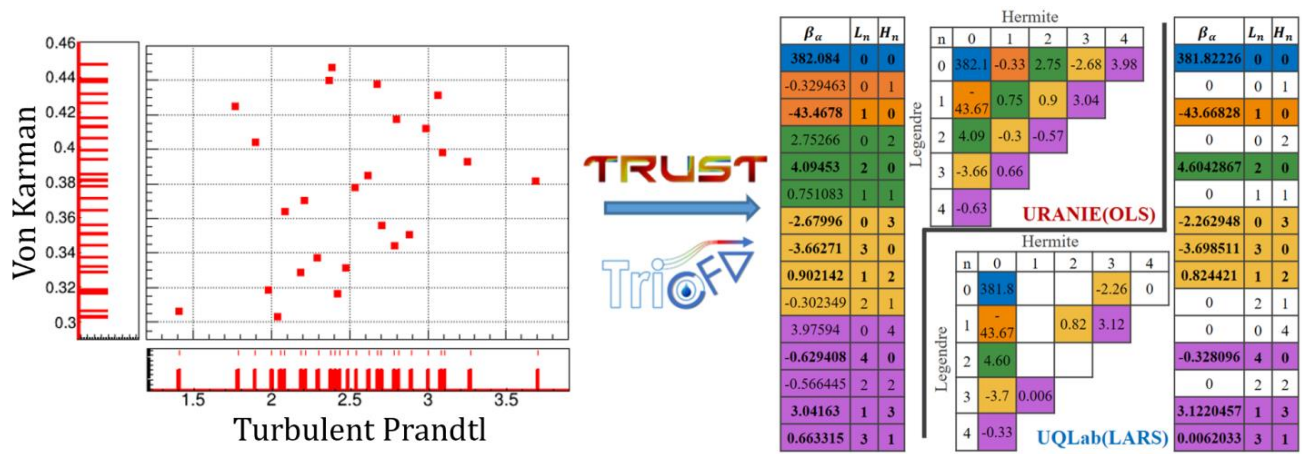

**Figure 7: DOE by using LHS and construction of PCE coefficients by Uranie and UQLab.**

**Table 4: Validation of meta-model for the mass flux by PCE on Uranie and UQLab.**

| Platform | Mean value | Std Dev. | Order | Coefficient | $1 - R^2$ | $Q^2$ | Valid. Err |
|---|---|---|---|---|---|---|---|
| Uranie | 382.08 | 44.30 | 4 | 15 | 4.05E-04 | 0.9406 | 9.3E-02 |
| UQLab | 381.82 | 44.24 | 4 | 9 | 6.09 E-04 | 0.9987 | 5.6E-03 |

---

[2] https://sourceforge.net/projects/uranie/ with Ordinary Least Squares Regression (OLS) method for PCE
[3] https://www.uqlab.com/ with Least Angle RegreSsion (LARS) method for PCE

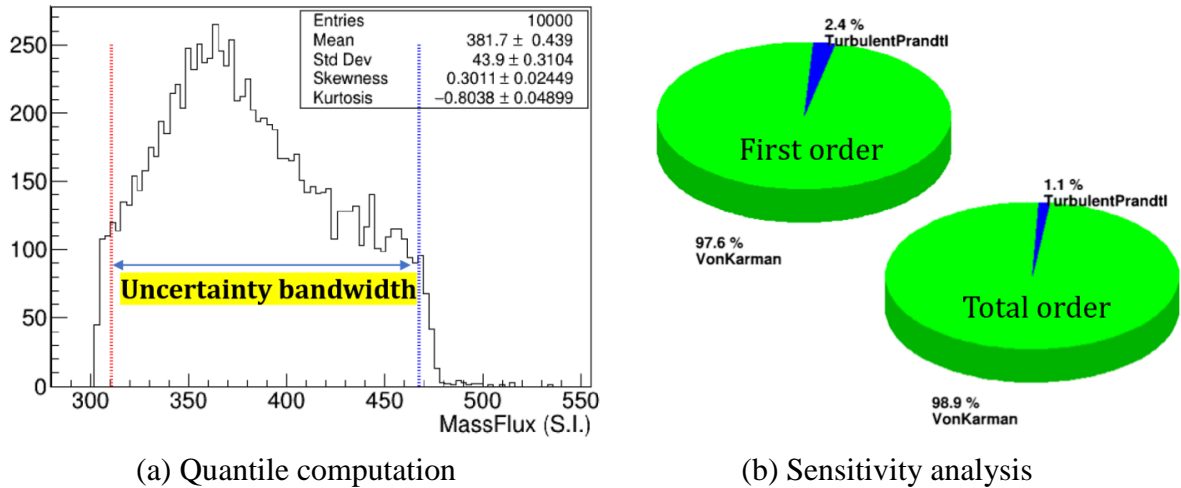

(a) Quantile computation  (b) Sensitivity analysis
**Figure 8: Quantile computation and sensitivity analysis on mass-flux for LEVM.**

Subsequent quantile computation and calculation of the uncertainty bandwidth are significantly influenced by the probabilistic distributions of input variables. We observe a wide range of uncertainty bandwidth for LEVM in Figure 8a. However, for LES calculations of a similar nature, the bandwidth range remains within 4%. Then, sensitivity analysis helps identify the most influential parameters, reaffirming the dominance of the wall function in Figure 8b. Similarly, the sub-grid coefficient in LES also dominates over turbulent Prandtl number. Finally, the calibration analysis presented in Table 5 demonstrates useful in enhancing lower-fidelity simulations, specifically by calibrating closure laws (wall function here) against high-fidelity (reference) simulations. This initial calibration step ensures that lower-fidelity models, once adjusted, serve effectively in meta-modelling due to their efficiency in predicting the quantity of interest. Von Karman coefficient present in wall functions with a standard value of 0.415 for pressure driven flows [12], is calibrated for better accuracy in natural convection scenarios (value of 0.367). As already mentioned, the reliability of this calibration for reactor case is a central upscaling question.

**Table 5: Calibration of LEVM using Approximate Bayesian Computation (ABC) method.**

| Coefficient | Mean value | Std deviation |
|---|---|---|
| Von Karman (in a generalized wall function) | 0.367 ($<$ 0.415) | 0.0062 |
| Turbulent Prandtl (in energy equation) | (0.9$<$)1.476 | 0.172 |

## 4. SCALING INVESTIGATION BY HIGHER-FIDELITY SIMULATIONS

Scaling investigations in simulations encompass various aspects like geometry, fluid mechanics, heat transfer, discretization, and closure laws. For the different fidelity simulations outlined in Table 1, several steps require scaling investigation. For two types of uncertainties quantified, we will define the Scales-of-Interest (SoIs) for our target scenario. Subsequently, we examine the scale effects by focusing on mass flow at steady state as a Figure-of-Merit (FoM).

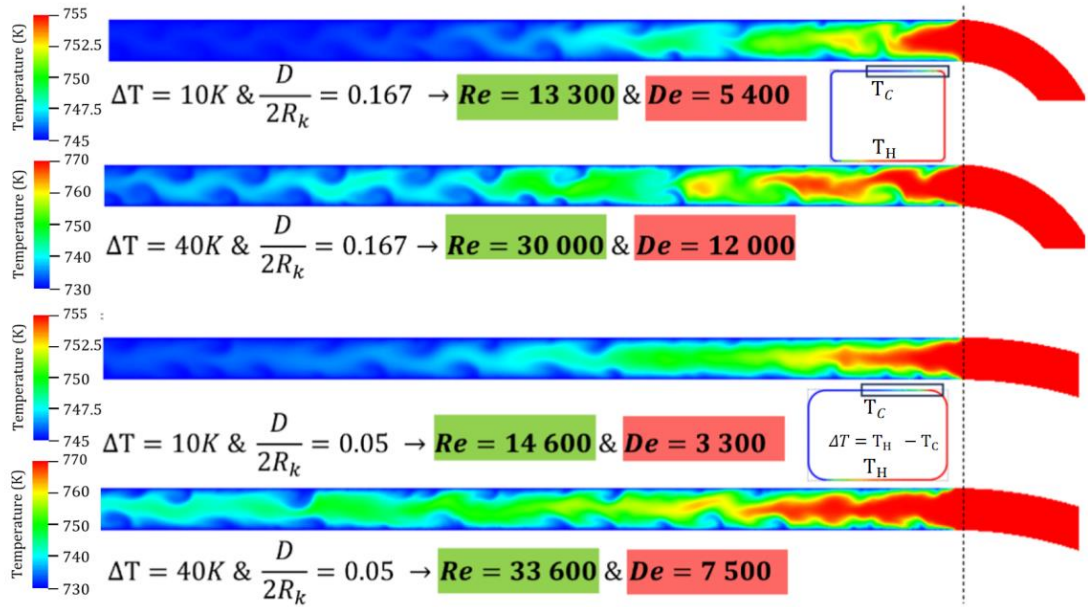

**Figure 9: NCL flow establishment for different input power (imposed by temperature difference) and curvature ratio by LES WALE simulations.**

### 4.1 Identification of Scale-of-Interest (SoI)

As a reminder, the modelling process is impacted by both numerical and physical distortions, influenced by the chosen mesh size and the physical stiffness considered. In transitioning from one physical scale to another during the NC establishment, we define several physical SoI: Reynolds, Rayleigh and Dean numbers, the latter being based on centrifugal effects. Some indicative values are presented in Figure 9 where the sharper elbows (elbow curvature ratio of 0.167) generate more local recirculation and larger Dean number. To make the scale effect easier to notice, various configurations with different dimensionless ratios (Table 6) are examined for different elbow radius $R_k$ and pipe diameters D. Particularly, focusing on the elbow scale effect, six levels of physical stiffness are investigated while other dimensionless ratios remain constant.

**Table 6: Configuration of HHHC loops with different dimensionless geometry ratios.**

| Scale | $S_1$ | $S_2$ | $S_3$ | $S_4$ | $S_5$ | $S_6$ |
|---|---|---|---|---|---|---|
| D | 25 | 25 | 25 | 25 | 25 | 25 |
| $R_k$/D | 3 | 4 | 5 | 6 | 7.2 | 10 |
| D/2$R_k$ | 0.167 | 0.125 | 0.1 | 0.083 | 0.069 | 0.05 |
| $L_H$/D | 60 | 60 | 60 | 60 | 60 | 60 |
| H/D | 52 | 52 | 52 | 52 | 52 | 52 |

### 4.2 Scale effect on modelling uncertainties by HF simulations

#### 4.2.1 Numerical uncertainties for discretization under scale change

Our previous study [13] revealed scaling effects on numerical uncertainty quantified by GCI and uncertainty propagation using Monte Carlo sampling. For the NC onset scenario, we proposed new invariants for critical Rayleigh number and its uncertainty bandwidth, the latter exhibiting good scalability when physical distortions are adequately addressed. Although NC onset involves low Reynolds and transitional regime, final regime involves turbulent flow.

When the input temperature difference between heat source and sink ΔT is changed from 10K to 40K (Figure 10a) we observe an increase in the uncertainty bands correlated with rising Reynolds numbers. Conversely, uncertainties for flow rate decrease with increasing Dean number (increasing $D/2R_k$ ratio). Thus, while anisotropic turbulence can be seen to increase jointly with Reynolds and Dean numbers, this is not the case for numerical uncertainties, highlighting the challenge in correlating numerical uncertainties with different SoIs (different distortions).

### 4.2.2 *Physical modelling uncertainties for input parameters under scale change*

To continue with the uncertainties of the physical model, we performed hundreds of LES using its prior meta-modelling (following the same approach as presented for URANS in section 3.2), in order to determine the uncertainty bands. Comparing uncertainties across different values of SoI, Figure 10b exhibits increasing of the uncertainty bands with rising Reynolds numbers, similarly to above findings for numerical aspects. However, uncertainty bands change only slightly over different Dean scales. The resolved eddy ratio can be quantitatively correlated with the slight degradation in physical modelling uncertainty [11]. In the end, through NC establishment, physical modelling uncertainties evolve also with given SoIs, but less dramatic than numerical ones, underlying the general scalability of such a high-fidelity simulation and the loosely interest of correlating an uncertainty change with distortion effect. Conversely, the LF approach, the only one applicable at the reactor scale, is likely to present a reduced predictability. If so, the HF approach should be effective in pinpointing the LF weaknesses: the concept of scaling uncertainty could be introduced as the growth of the LF simulation uncertainty associated with distortion effects. In line, a specific methodology within the BEPU framework - leveraging both HF and LF approaches - could pragmatically enable correlating distortion effects with scaling uncertainty, thereby providing a metric principle.

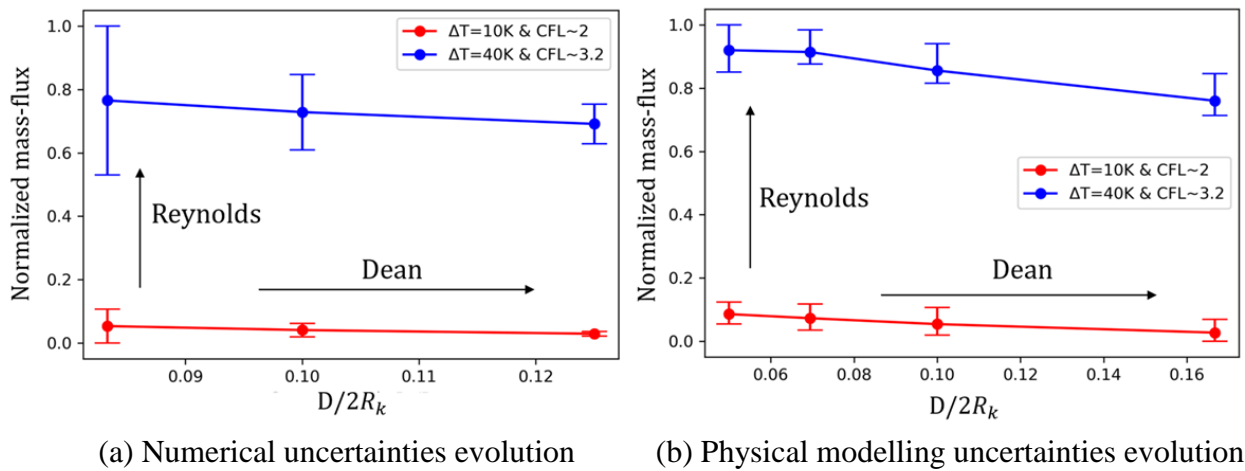

(a) Numerical uncertainties evolution     (b) Physical modelling uncertainties evolution

**Figure 10: Scaling effect of modelling uncertainties of normalized mass flow-rate for different curvature ratios.**

## 5. CONCLUSION

New nuclear plants designs rely on passive NC mechanism for the residual decay heat removal. In the BEPU framework, TrioCFD simulations are carried out using liquid sodium as the coolant in a HHHC loop, which is regarded as a singular configuration in the literature. Both High-Fidelity (HF via LES) and Lower-Fidelity (LF via URANS) approaches are used, thus highlighting differing levels of turbulence and boundary layer modelling. In this work, some specific uncertainties are quantified: they embrace both numerical solution uncertainties, tackled by the Grid Convergence Index, and physical modelling uncertainties, managed through Polynomial Chaos Expansion via the URANIE platform.

This study is focused on the uncertainty evolution regarding the physical distortion effects, sketching the notion of scaling uncertainties. From the cases computed using LES, it comes out that only the numerical solution uncertainties are significantly impacted by the physical distortion. This underlines the high predictive solution of the LES approach. It can be expected that URANS studies with wall laws (featuring reasonably CPU-efficient approach for a reactor-scale application) would experience a degradation in predictability due to distortion effects. Indeed, velocity and temperature profiles in the NC boundary layers of liquid sodium given by LES show their specificity compared with the standard laws from forced convection flows.

Hence, this study suggests that a specific methodology within the BEPU framework - leveraging both HF and LF approaches - could pragmatically enable correlating distortion effects with scaling uncertainty, thereby providing a metric principle. The latter is aligned with the request of ASN Guide 28, which requires proof of a tool's predictability. Therefore, the focus should loosely be on enlarging input parameter uncertainties of closure laws to cover two (possibly numerical) experiments featuring scale effects, but rather on measuring the change in calibration uncertainties required to shift from one experiment to another. HF could provide some reference data to do so and provide detailed insights to anticipate risk evaluation while applying a lower fidelity approach.

In a bigger picture, the Dynamical System Scaling technique may be diverted from its original purpose (that is reducing distortion) by instead offering original suggestions of well-controlled distorted configurations that could be included in the validation perimeter. Main thoughts discussed in this paper are being integrated in a global methodology MUSQ (Modelling Uncertainties with Scaling Quantification) [11].

### ACKNOWLEDGMENTS

Thanks to the RNR-Na project within the 4[th] Generation Program of the CEA in the context of which this work took place.